\documentclass{article} 
\usepackage{epsfig} 
\usepackage{cite} 
\def\ln{\hbox{ln}}

\begin{document} 
\unitlength1cm 
\begin{titlepage} 
\vspace*{-1cm} 
\begin{flushright} 
TTP02-39\\ 
November 2002 
\end{flushright} 
\vskip 3.5cm 
\renewcommand{\topfraction}{0.9} 
\renewcommand{\textfraction}{0.0} 

\begin{center} 
\boldmath 
{\Large\bf The analytic value of a 3-loop sunrise graph 
in a particular kinematical configuration. 
}\unboldmath 
\vskip 1.cm 
{\large P. Mastrolia}$^{a,b}$ and 
{\large E. Remiddi}$^{a,c}$ 
\vskip .7cm 
\vskip .4cm 
{\it $^a$ Dipartimento di Fisica, 
    Universit\`{a} di Bologna, I-40126 Bologna, Italy \\
       $^b$ Institut f\"ur Theoretische Teilchenphysik,
            Universit\"at Karlsruhe, D-76128 Karlsruhe, Germany \\ 
       $^c$ INFN, Sezione di Bologna, I-40126 Bologna, Italy \\ 
} 
\end{center} 
\vskip 2.6cm 
\begin{abstract} 
We consider the scalar integral associated to the 3-loop sunrise graph with 
a massless line, two massive lines of equal mass $M$, a fourth line of 
mass equal to $Mx$, and the external invariant timelike and equal to the 
square of the fourth mass. We write the differential equation in $x$ 
satisfied by the integral, expand it in the continuous dimension $d$ 
around $d=4$ and solve the system of the resulting chained differential 
equations in closed analytic form, expressing the solutions in terms of 
Harmonic Polylogarithms. As a byproduct, we give the limiting values of the 
coefficients of the $(d-4)$ expansion at $x=1$ and $x=0$. 
\end{abstract} 
\vfill 
\end{titlepage} 

\newpage 
\renewcommand{\theequation}{\mbox{\arabic{section}.\arabic{equation}}} 
\section{Introduction} 
\label{sec:int} 
\begin{figure}[h] 
\begin{center} 
\includegraphics*[2cm,1cm][14cm,6cm]{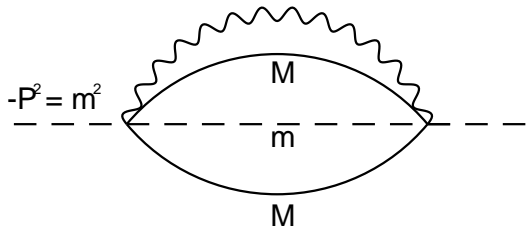} 
\caption{The considered 3-loop sunrise graph.} 
\end{center}
\label{fig:1} 
\end{figure}

\setcounter{equation}{0} 
This paper is devoted to the analytic evaluation of the scalar integral 
associated to the 3-loop 
sunrise graph with a massless line, two massive lines of equal mass $M$, 
a fourth line of mass $m$, with $m \ne M$, and the external invariant 
timelike and equal to $m^2$, as depicted in Fig.~\ref{fig:1}. 
We use the same method as in ~\cite{sun2}, which deals however with 
a 2-loop case. We start by writing the differential equation in 
$x=m/M$ satisfied by the considered integral (which turns out to 
be a third order differential equation), expand it in $(d-4)$, where 
$d$ is the continuous dimension which regularizes the integral, and 
solve the resulting system of chained differential equations for the 
coefficients of the expansion in $(d-4)$ by repeated use of the 
variation of constants formula by Euler. The approach allows to obtain 
the coefficients of the expansion of the integral in $(d-4)$ up to 
virtually any order in $(d-4)$, expressing all of them in terms of 
Harmonic Polylogarithms (HPL's~\cite{hpl}) of argument $x$. 
At $x=0,1$ we recover the two 
quantities $J_{13}$ and $J_{11}$ introduced in~\cite{Krakow} for the 
analytic evaluation of the $(g-2)$ of the electron at 3 loops in QED.
$J_{13}$, which corresponds to the vacuum graph of Fig.2, and $J_{11}$, 
which is the on-mass-shell value of the graph of Fig.1 at $x=1$, 
where evaluated analytically in~\cite{Krakow} up to order $(d-4)^2$ 
and $(d-4)^3$ respectively. As a byproduct of the analytic result 
established for arbitrary $x$, by taking its limiting values at $x=0$ 
and $x=1$ we obtain two more terms of the expansion in $(d-4)$ of 
$J_{13}$ (the analytic value of $J_{13}$, exact in $d$, can be easily 
obtained by direct integration of the corresponding loop integral, 
so that this result can be seen as a check of the approach) and of 
$J_{11}$ (a result not yet known in the literature). 
Those higher order terms in the $(d-4)$ expansion are expected 
to be needed in the analytic evaluation of 4-loop QED static quantities 
(such as $(g-2)$ of the electron, charge slope and renormalization 
counterterms). 

\section{The differential equation} 
\label{sec:Thediffeq} 
\setcounter{equation}{0} 
Our aim is to evaluate the scalar integral 
\begin{equation} 
  F(d,M^2,m^2,P^2=-m^2) = \frac{1}{(2\pi)^{3(d-2)}} 
        \int \frac{d^dk_1\  d^dk_2 \ d^dk_3 } 
      { k_1^2 (k_2^2+M^2) (k_3^2+M^2) [(P-k_1-k_2-k_3)^2+m^2] } \ .
\label{eq:defF} 
\end{equation} 
In terms of the dimensionless variable $ x = m/M $ and putting 
$P=Mp$ one can define the function $\Phi(d,x)$ as 
\begin{equation} 
     F(d,M^2,m^2,P^2=-m^2) = M^{(3d-8)} \ C^3(d)\ \Phi(d,x) \ , 
\label{eq:FtoPhi} 
\end{equation} 
where $C(d)= (4\pi)^\frac{4-d}{2}\Gamma(3-d/2) $ is an overall normalization 
factor, with the limiting value $C(4)=1$ at $d=4$. The previous equation 
can also be written as 
\begin{equation} 
  \Phi(d,x) = \frac{C^{-3}(d)}{(2\pi)^{3(d-2)}} \int 
     \frac{d^dk_1\  d^dk_2 \ d^dk_3 } 
     { k_1^2 (k_2^2+1) (k_3^2+1) [(p-k_1-k_2-k_3)^2+x^2] } \ , 
                                          \quad (p^2=-x^2) \ . 
\label{eq:defPhi} 
\end{equation} 

Following~\cite{sun2}, for evaluating Eq.(\ref{eq:defPhi}) we will 
first establish the differential equation in $x$ satisfied by 
$ \Phi(d,x) $. For writing the differential equation, we have to find the 
{\it Master Integrals} (MI's) of the graph of Fig.1 . That is done by 
solving the {\it Integration by Parts Identities} (IBP-id's)~\cite{IBP}, 
for the integrals associated with the graph of Fig. 1.
We generated an initial system of 492 equations involving 449 unknown 
associated integrals. By using the symbolical solutions~\cite{bastei} of 
the IBP-id's it becomes a system of 205 equations in 76 unknown integrals. 
By solving explicitly that last system, we were left with four MI's only, 
which we take to be the fully scalar integral of Eq.(\ref{eq:defF}) 
and three more integrals, having as integrand the same denominator as in 
Eq.(\ref{eq:defF}) and numerators respectively equal to the scalar products 
$(k_1\cdot k_2), (k_2\cdot k_3)$ and $(P\cdot k_2)$. 
Those four MI's satisfy a system of four $1^{st}$ Order 
Differential Equations in the $x$ variable. We rewrite them in terms 
of a single higher order equation for $\Phi(d,x)$. In so doing, it 
turns out that the fourth MI decouples, and $\Phi(d,x)$ is found to 
satisfy the following $3^{rd}$ Order Differential Equation in $x$, 
exact in $d$, 
\begin{eqnarray} 
 x^2 (1-x^2) \frac{d^3}{dx^3} \Phi(d,x) 
+ \biggl\{ \bigl[2 - (d-4)\bigr]  + \bigl[2 + 5 (d-4)\bigr] x^2 \biggr\} 
        x \frac{d^2}{dx^2} \Phi(d,x) &  & \nonumber \\ 
    + \biggl\{ \bigl[- 6 - 8(d-4) - 2(d-4)^2\bigr] 
             + \bigl[2 - 4(d-4) - 6(d-4)^2\bigr] x^2 \biggr\} 
       \frac{d}{dx} \Phi(d,x) &  & \nonumber \\ 
  - \bigl[8 + 14(d-4) + 6(d-4)^2\bigr] x \ \Phi(d,x) & = & 
    \frac{x\ x^{(d-4)}}{(d-4)^3} \ . \qquad \qquad 
\label{eq:Phidiffeq} 
\end{eqnarray} 
Let us note that the above equation looks slightly simpler when written 
in terms of $x^2$; but as $x$ is the proper variable for expressing 
the final results, we use $x$ since the very beginning. 

\section{The $x \to 0$ behaviour.} 
\label{sec:xto0} 
\setcounter{equation}{0} 
For studying the $x \to 0$ behaviour, we can try to solve 
Eq.~(\ref{eq:Phidiffeq}) with the ansatz 
\begin{equation} 
   \Phi(d,x) = \sum_i x^{\alpha_i} 
               \left( \sum_{n=0}^\infty a_n^{(i)}(d) \ x^{n} \right) \ . 
\label{eq:ansatz} 
\end{equation} 
Four possible values for the exponents $\alpha_i$ are found: 
\begin{eqnarray} 
  \alpha_1 &=& 0 \ , \nonumber\\ 
  \alpha_2 &=& -(d-2) \ , \nonumber\\ 
  \alpha_3 &=& (2d - 5) \ , \nonumber\\ 
  \alpha_4 &=& (d-2) \ , 
\label{eq:alphas} 
\end{eqnarray} 
where the last value is forced by the inhomogeneous term, while the 
first three behaviours are those of the solutions of the homogeneous 
part of the equation. By inspection of Eq.(\ref{eq:defPhi}), we see 
that the aimed $\Phi(d,x)$ is finite in the  $x\to 0$ limit when $d > 2$. 
That is true even when the continuous parameter $d$ is in the range 
$ 2 < d < 2.5 $; as in that range of $d$ the exponents 
$ \alpha_2, \alpha_3 $ give terms which diverge for $ x\to 0$, the 
corresponding behaviours are ruled out. In the $d\to4$ limit, that 
implies the absence of terms which for $x\to 0$ diverge as $1/x^2$ 
(corresponding to $ \alpha_2 $) or behave as odd powers of $x$ 
(corresponding to $ \alpha_3 $). 
That gives two conditions which can be used for determining two of the 
three arbitrary constants present in the solution of Eq.(\ref{eq:Phidiffeq}). 
Further, only even powers of $x$ are found to appear inside the 
brackets of Eq.(\ref{eq:ansatz}), so that the 
actual expansion of $\Phi(d,x)$ for $x\to 0$ has the form 
\begin{eqnarray} 
  \Phi(d,x) &=&   \sum_{k=0}^{\infty} a_{2k}(d)x^{2k} \nonumber \\ 
            &+& x^{(d-2)} \sum_{k=0}^{\infty} b_{2k}(d)x^{2k} \ . 
\label{eq:Phiat0} 
\end{eqnarray} 
The equation determines the coefficients $ a_{2k}(d) $ in terms of the 
first coefficient $ a_0(d) $ (which must then fixed by some different 
method), while all the coefficients $ b_{2k}(d) $ are fully fixed 
by the inhomogeneous term of the equation. Explicitly, one finds 
\begin{eqnarray} 
 a_2(d) & = & - \ \frac{(3 d - 8) (d - 3)}{d \ (2 d - 7)} \ a_0(d)
 \ , \nonumber \\ 
 &  & \nonumber \\ 
 a_4(d) & = & \frac{3}{2} \ 
           \frac{(3 d - 10) (3 d - 8) (d - 3) (d - 4)} 
                {d \ (d + 2) (2 d - 7) (2 d - 9)} \ a_0(d) 
 \ , \nonumber \\ 
 &  & \\ 
 b_0(d) & = & - \frac{1}{2}\ \frac{1}{(d - 2)^2 (d - 3) (d - 4)^3}
 \ , \nonumber \\ 
 &  & \nonumber \\ 
 b_2(d) & = & \frac{1}{2}\ \frac{1}{d \ (d - 2)^2 (d - 4)^2 (d - 5)} 
 \ , \nonumber 
\end{eqnarray}
etc. 

\section{The $x \to 1$ behaviour.} 
\label{sec:xto1} 
\setcounter{equation}{0} 
A similar discussion applies to the $x\to 1$ behaviour. 
When solving Eq.~(\ref{eq:Phidiffeq}) with the ansatz 
$$ 
   \Phi(d,x) = \sum_i (1-x)^{\beta_i} 
               \left( \sum_{n=0}^\infty c_n^{(i)}(d) \ (1-x)^{n} \right) \ , 
$$ 
one finds for the exponents $\beta_i$ the values 
\begin{eqnarray} 
  \beta_1 &=& 0 \ , \nonumber\\ 
  \beta_2 &=& 1 \ , \nonumber\\ 
  \beta_3 &=& 2(d - 2) \ , \nonumber\\ 
  \beta_4 &=& 2 \ , 
\label{eq:betas} 
\end{eqnarray} 
the first three due to the homogeneous part of the equation, the 
fourth to the inhomogeneous term. By inspection of Eq.(\ref{eq:defPhi}), 
one finds that the integral is regular at $x=1$, so that only the 
behaviour corresponding to $\beta_3$ is ruled out. When expanded in 
powers of $(d-4)$ in the $d\to4$ limit, it would generate 
for $x\to 1$ terms in $\ln(1-x)$, which are therefore forbidden. 
That information, coming from the study of the behaviour at $x=1$, 
provides with a further condition for the determination of the three 
integration constants. 

\section{The expansion in $(d-4)$.} 
\label{sec:expind} 
\setcounter{equation}{0} 
The 3-loop integral $\Phi(d,x)$, Eq.(\ref{eq:defPhi}), develops a triple 
pole in $(d-4)$ when $d\to4$, so that its Laurent's expansion reads 
\begin{equation} 
 \Phi(x,d) = \sum_{n=-3}^\infty (d-4)^n \ \Phi^{(n)}(x) \ . 
\label{eq:Laurent}
\end{equation} 
By inserting it into Eq.(\ref{eq:Phidiffeq}), the coefficients 
$ \Phi^{(n)}(x) $ are found to satisfy the following system of chained 
differential equations 
\begin{eqnarray} 
 &   & 
\left[ 
   \frac{d^3}{dx^3} 
 + 2 \left(   \frac{1}{x} + \frac{1}{1-x} 
            - \frac{1}{1+x} 
     \right) \frac{d^2}{dx^2} 
 - 2 \left( \frac{3}{x^2} + \frac{1}{1-x} 
            + \frac{1}{1+x} 
     \right) \frac{d}{dx} 
 - 4 \left( \frac{2}{x} + \frac{1}{1-x} 
            - \frac{1}{1+x} 
     \right) 
\right] \Phi^{(n)}(x) \nonumber \\ 
 &   &  \nonumber \\ 
 & = & \left( \frac{1}{x} - \frac{2}{1-x} + \frac{2}{1+x} 
       \right) \frac{d^2}{dx^2}\Phi^{(n-1)}(x) \nonumber \\ 
 &   &  \nonumber \\ 
 &   & 
       + 2 \left( \frac{4}{x^2} + \frac{3}{1-x} + \frac{3}{1+x} 
           \right) \frac{d}{dx}\Phi^{(n-1)}(x) 
       + 7 \left( \frac{2}{x} + \frac{1}{1-x} - \frac{1}{1+x} 
           \right) \Phi^{(n-1)}(x) \nonumber \\ 
 &   &  \nonumber \\ 
 &   & + 2 \left( \frac{1}{x^2} + \frac{2}{1-x} + \frac{2}{1+x} 
           \right) \frac{d}{dx}\Phi^{(n-2)}(x) 
       + 3 \left( \frac{2}{x} + \frac{1}{1-x} - \frac{1}{1+x} 
           \right) \Phi^{(n-2)}(x) \nonumber \\ 
 &   &  \nonumber \\ 
 &   & + \frac{1}{2} \left( \frac{2}{x} + \frac{1}{1-x} - \frac{1}{1+x} 
           \right) \frac{1}{(n+3)!} \ \ln^{(n+3)}(x) \quad , 
\label{eq:Phindiffeq} 
\end{eqnarray}
where $\Phi^{(n)}(x) = 0$ when $n < -3$. Note that in the equation for 
$\Phi^{(-3)}(x)$ the inhomogeneous term is just the proper term of the 
expansion in $(d-4)$ of the inhomogeneous term of Eq.(\ref{eq:Phidiffeq}), 
which is known (as matter of fact, a rational factor), 
while in the equation for $\Phi^{(-2)}(x)$, $\Phi^{(-3)}(x)$ appears in the 
{\it r.h.s.} as a contribution to the inhomogeneous term and so on; 
given such a chained structure of the equations, it is natural to solve 
them recursively, starting from the equation for $\Phi^{(-3)}(x)$ 
and going up to the successive coefficients of the expansion in $(d-4)$ 
of $\Phi(d,x)$. 
Summarizing, all the equations (\ref{eq:Phindiffeq}) have the same form, 
\begin{eqnarray} 
\left[ 
   \frac{d^3}{dx^3} 
 + 2 \left(   \frac{1}{x} + \frac{1}{1-x} 
            - \frac{1}{1+x} 
     \right) \frac{d^2}{dx^2} 
 - 2 \left( \frac{3}{x^2} + \frac{1}{1-x} 
            + \frac{1}{1+x} 
     \right) \frac{d}{dx} 
 - 4 \left( \frac{2}{x} + \frac{1}{1-x} 
            - \frac{1}{1+x} 
     \right) 
\right] \Phi^{(n)}(x) & = & \nonumber \\
 K^{(n)}(x) \ , & &  
\label{eq:diffeqn} 
\end{eqnarray} 
where $K^{(n)}(x)$ stands for the inhomogeneous term, which can be taken 
as known when solving the equations starting from $n=-3$ and continuing 
in order of growing $n$. 

\section{The solution.} 
\label{sec:solution} 
\setcounter{equation}{0} 
The homogenous equation associated to all the equations corresponding 
to different values of $n$ is always the same, namely 
\begin{equation} 
\left[ 
   \frac{d^3}{dx^3} 
 + 2 \left(   \frac{1}{x} + \frac{1}{1-x} 
            - \frac{1}{1+x} 
     \right) \frac{d^2}{dx^2} 
 - 2 \left( \frac{3}{x^2} + \frac{1}{1-x} 
            + \frac{1}{1+x} 
     \right) \frac{d}{dx} 
 - 4 \left( \frac{2}{x} + \frac{1}{1-x} 
            - \frac{1}{1+x} 
     \right) 
\right] \phi(x) = 0 \ , 
\label{eq:homodiffeq} 
\end{equation} 
so that all the Eq.s(\ref{eq:diffeqn}) can be solved by Euler's method 
of the variation of the constants. In this case the general solution can 
be written as 
\begin{eqnarray}
  \Phi^{(n)}(x) &=& + \phi_1(x) \left[ \Phi^{(n)}_1
           + \int^x\;dy \frac{\phi_2(y)\phi_3^{'}(y)
           - \phi_2^{'}(y)\phi_3(y)}{ W(y) } \ K^{(n)}(y) \right] \nonumber\\
          & & + \phi_2(x) \left[ \Phi^{(n)}_2
           - \int^x\;dy \frac{\phi_1(y)\phi_3^{'}(y)
           - \phi_1^{'}(y)\phi_3(y)}{ W(y) } \ K^{(n)}(y) \right] \nonumber\\
          & & + \phi_3(x) \left[ \Phi^{(n)}_3
           + \int^x\;dy \frac{\phi_1(y)\phi_2^{'}(y)
           - \phi_1^{'}(y)\phi_2(y)}{ W(y) } \ K^{(n)}(y) \right] \ ,
\label{eq:Euler} 
\end{eqnarray} 
where the $\phi_i(x), i=1,2,3$ are three linearly independent solutions 
of the homogeneous equation Eq.(\ref{eq:homodiffeq}), $ W(x) $ their 
Wronskian and the $\Phi^{(n)}_i, i=1,2,3$ the integration constants. 
\par 
Although algorithms for solving differential equations, even 
with simple rational coefficients, are not available in general, by 
trial and error it is not difficult to find (and in any case easy to 
verify) that Eq.(\ref{eq:homodiffeq}) admits the three linearly 
independent solutions 
\begin{eqnarray}
\phi_1(x) & = & (1 - x^2) \ ,  \nonumber\\ 
\phi_2(x) & = & - \ \frac{1}{2} \frac{(1 - x^2) (1 - x^4) }{ x^2 } 
                - 2 (1 - x^2) H(0;x)  \ , \nonumber\\ 
\phi_3(x) & = & + \ \frac{3}{512} \frac{(1 - x^2) (1 - x^4) }{ x^2 }
                  \ [H(-1;x) + H(1;x)] \nonumber \\
          &   & + \frac{3}{128} (1 - x^2) \ [H(0,-1;x) + H(0,1;x)]
                  \nonumber \\
          &   & - \ \frac{1}{256} \frac{(x^2 + 3) (3 x^2 + 1)}{x} \ , 
\label{eq:homosol} 
\end{eqnarray} 
where the functions $H(0;x), H(-1;x), H(1;x), H(0,-1;x)$ etc. are 
harmonic polylogarithms~\cite{sun2},~\cite{hpl}. The Wronskian of the 
three above functions $\phi_i(x)$ is 
\begin{equation}
 W(x) = - \frac{(1 - x)^2 (1 + x)^2}{x^2} \ .
\label{eq:Wronskian} 
\end{equation} 
When inserting Eq.s(\ref{eq:homosol}), Eq.(\ref{eq:Wronskian}) and 
$K^{(n)}(x)$, as defined by Eq.s(\ref{eq:Phindiffeq},\ref{eq:diffeqn}), 
into Eq.(\ref{eq:Euler}), one sees that in analogy to previous 
work~\cite{sun2} all the integrations can be carried out in 
terms of harmonic polylogarithms, up to any desired value of $n$, 
and one is left only with the problem of fixing, for any order $n$, 
the three integration constants $\Phi^{(n)}_i$. \par 
It is worth emphasizing that the \emph{quantitative} determination of 
the integration constants can be obtained by simply imposing to the 
solution the \emph{qualitative} behaviours for $x\to 0$, $x\to 1$ 
discussed in the previous sections Sec.(\ref{sec:xto0},\ref{sec:xto1}). 
More exactly, by imposing the right qualitative behaviour at order $n$, 
one obtains the explicite, quantitative value of the integration 
constants up to order $n-2$ included (in other words: imposing the 
right qualitative behaviour to $\Phi^{(-1)}(x)$ one fixes the 
first three constants $\Phi^{(-3)}_i$, and so on). We have evaluated 
the solutions up to $n=5$ (the results involves harmonic polylogarithms 
of weight 8), and fixed the integration constants 
up to $n=3$ included. The integration constants were found to be 
$$ \Phi^{(k)}_3 = 0 \hskip 5mm {\mbox{ (for any $k$)}} \ ; $$ 
\begin{eqnarray} 
\Phi^{(-3)}_1 & = & \frac{7}{192} \ ;  \hskip 5mm 
\Phi^{(-3)}_2 =  - \frac{3}{128} \ ;  \hskip 5mm 
\Phi^{(-2)}_1  =  - \frac{19}{256} \ ;  \hskip 5mm 
\Phi^{(-2)}_2  =  \frac{37}{512} \ ;  \nonumber \\ 
\Phi^{(-1)}_1 & = & \frac{7}{96} \ ;  \hskip 5mm 
\Phi^{(-1)}_2  =  - \frac{233}{3072} \ ;  \hskip 5mm 
\Phi^{(0)}_1  =  - \frac{259}{4608} 
                   + \frac{1}{24} \zeta(3)  \ ;  \hskip 5mm 
\Phi^{(0)}_2  =   \frac{1645}{36864} \ ;  \nonumber \\
\Phi^{(1)}_1 & = &    \frac{3395}{55296}
                    - \frac{13}{128} \zeta(3)
                    + \frac{1}{960} \pi^4\ ;  \hskip 5mm 
\Phi^{(1)}_2  =   - \frac{15149}{442368}
                    + \frac{9}{256} \zeta(3) \ ;  \nonumber \\
\Phi^{(2)}_1 & = &  - \frac{108691}{663552}
                    + \frac{97}{768} \zeta(3)
                    - \frac{13}{5120} \pi^4
                    + \frac{3}{16}\zeta(5) \ ;  \hskip 5mm 
\Phi^{(2)}_2  =     \frac{1038085}{5308416} 
                    - \frac{95}{1536} \zeta(3)
                    + \frac{9}{10240} \pi^4 \ ;  \nonumber \\
\Phi^{(3)}_1 & = &    \frac{4181051}{7962624}
                    - \frac{1225}{9216} \zeta(3)
                    + \frac{97}{30720} \pi^4
                    - \frac{117}{256} \zeta(5)
                    - \frac{1}{48} \zeta^2(3)
                    + \frac{1}{2688} \pi^6  \ ;  \nonumber \\
\Phi^{(3)}_2 & = &  - \frac{56294669}{63700992}
                    - \frac{19}{12288} \pi^4
                    + \frac{2833}{18432} \zeta(3)
                    + \frac{81}{512} \zeta(5) \ .  \nonumber 
\label{eq:intconst} 
\end{eqnarray} 
The explicit expressions of all the $\Phi^{(n)}(x)$ which we have 
evaluated are too long to be listed here, and we give only the values up 
to $n=0$: 
\begin{eqnarray}
 \Phi^{(-3)}(x) &=&  - \frac{1}{24} - \frac{1}{12} x^2 \ ; \\
 \Phi^{(-2)}(x) &=& + \frac{7}{96} + \frac{5}{32} x^2 - \frac{1}{96} x^4
       - \frac{1}{8} x^2  \ H(0;x)\ ; \ \\
 \Phi^{(-1)}(x) &=& - \frac{25}{384}  
       - \frac{71}{384} x^2 + \frac{35}{1152} x^4 
       + \frac{1}{4} x^2 \left( 1 - \frac{1}{8} x^2 \right) \ H(0;x) 
       - \frac{1}{16} x^2 \ H(0;x) H(0;x) \ ; \\
 \Phi^{(0)}(x) &=& + \frac{19}{1536} + \frac{107}{512} x^2 
                   - \frac{559}{13824} x^4 \nonumber\\ 
   & &    
       - \frac{1}{32} \left(  1 + \frac{25}{2} x^2 - \frac{35}{12} x^4
         \right) \ H(0;x) 
       + \frac{1}{8} x^2 \left( 1 - \frac{1}{8} x^2
                         \right) \ H(0;x) H(0;x) \nonumber \\
   & & 
       - \frac{1}{32} \frac{(1 - x^2) (1-x^4)}{x^2}
           [H(-1;x) H(0;x) - H(0;x) H(1;x) + H(0,1;x) - H(0,-1;x)]
         \nonumber \\
   & &
       - \frac{1}{48} x^2  \ H(0;x) H(0;x) H(0;x) 
       + \frac{1}{8}(1 -  x^2)  [H(0,1;x) H(0;x) - H(0,-1;x) H(0;x)] 
        \nonumber\\ 
   & &
       + \frac{1}{4}(1 -  x^2) 
         \left[ H(0,0,-1;x) - H(0,0,1;x) + \frac{1}{6}\zeta(3) \right] 
         \ .
\end{eqnarray} 
Having the explicit form of the solution for any $x$ it is almost trivial 
to obtain its value at $x=1$ and $x=0$. 
It is to be observed, in this respect, that the integration constants 
$\Phi^{(n-1)}_k, \Phi^{(n)}_k$ do not contribute to the $x=1$ value 
up to order $n$, so that with the explicit value of the integration 
constants up to $n=3$ included we can evaluate $\Phi(d,x=1)$ in
Fig.(2) up to order $n=5$ included. 
\begin{figure}[h] 
\begin{center} 
\includegraphics*[2cm,1.5cm][14cm,5.5cm]{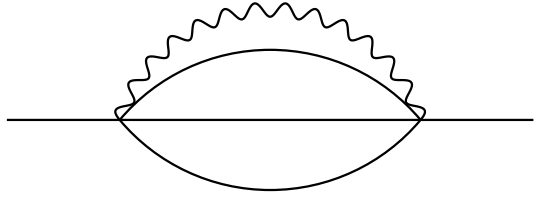} 
\caption{The considered 3-loop sunrise graph, in the case of equal
         masses, on mass-shell.} 
\end{center}
\label{fig:two} 
\end{figure} 
By using the tables of~\cite{VermTab} we find the 
following expansion in $(d-4)$ 
\begin{eqnarray}
 \Phi(d,x=1) & \stackrel{d \to 4}{=} & \nonumber \\
   & & \hspace*{-1.5cm}
       - \frac{1}{8} \frac{1}{(d-4)^3}
       + \frac{7}{32} \frac{1}{(d-4)^2} 
       - \frac{253}{1152} \frac{1}{(d-4)}  
       + \frac{2501}{13824} \nonumber \\
   & & \hspace*{-1.5cm}
       + (d-4)  \left(
          - \frac{59437}{165888} + \frac{1}{18} \pi^2
                \right) \nonumber \\
   & & \hspace*{-1.5cm}
       + (d-4)^2   \left(
          + \frac{2831381}{1990656}
          - \frac{71}{216} \pi^2
          + \frac{1}{3} \pi^2 \ln2
          - \frac{7}{9} \zeta(3)
          \right) \nonumber \\
   & & \hspace*{-1.5cm}
       + (d-4)^3   \left(
          - \frac{117529021}{23887872}
          + \frac{3115}{2592} \pi^2
          - \frac{71}{36} \pi^2 \ln2
          + \frac{497}{108} \zeta(3)
          + \frac{7}{9} \pi^2 \ln^2 2
          - \frac{43}{1080} \pi^4
          + \frac{2}{9} \ln^4 2
          + \frac{16}{3} a_4
          \right) \nonumber \\
   & & \hspace*{-1.5cm}
       + (d-4)^4   \left(
          + \frac{4081770917}{286654464}
          - \frac{109403}{31104} \pi^2  
          + \frac{3115}{432} \pi^2 \ln2
          - \frac{21805}{1296} \zeta(3)
          - \frac{497}{108} \pi^2 \ln^2 2
          + \frac{3053}{12960} \pi^4  \right. \nonumber \\
   & & \hspace*{-1.5cm}
         \qquad \qquad \quad \left.
          - \frac{71}{54} \ln^4 2    
          - \frac{284}{9} a_4  
          - \frac{43}{180} \pi^4 \ln2
          + \frac{14}{9} \pi^2 \ln^3 2
          + \frac{7}{36} \pi^2 \zeta(3)
          + \frac{4}{15} \ln^5 2
          + \frac{341}{12} \zeta(5)
          - 32 a_5
          \right) \nonumber \\
   & & \hspace*{-1.5cm}
       + (d-4)^5   \left(
          - \frac{125873914573}{3439853568}
          + \frac{3386467}{373248} \pi^2
          - \frac{109403}{5184} \pi^2 \ln2
          + \frac{765821}{15552} \zeta(3) 
          + \frac{21805}{1296} \pi^2 \ln^2 2
        \right. \nonumber \\
   & & \hspace*{-1.5cm}
         \qquad \qquad \quad \left.
          - \frac{26789}{31104} \pi^4  
          + \frac{3115}{648} \ln^4 2
          + \frac{3115}{27} a_4
          - \frac{497}{54} \pi^2 \ln^3 2  
          - \frac{497}{432} \pi^2 \zeta(3)
          + \frac{3053}{2160} \pi^4 \ln2
         \right. \nonumber \\
   & & \hspace*{-1.5cm}
         \qquad \qquad \quad \left.
          - \frac{24211}{144} \zeta(5)
          + \frac{568}{3} a_5
          - \frac{71}{45} \ln^5 2
          + \frac{41}{810} \pi^6
          + \frac{10}{9} \pi^2 \ln^4 2
          - \frac{151}{180} \pi^4 \ln^2 2
          + \frac{88}{3} \zeta(3) \ln^3 2 
         \right. \nonumber \\
   & & \hspace*{-1.5cm}
         \qquad \qquad \quad \left.
          - \frac{27}{2} \pi^2 \zeta(3) \ln2 
          + 176 \zeta(5) \ln2 
          - 176 a_5 \ln2 
          + \ln^6 2 
          + \frac{2723}{36} \zeta(3)^2 
          - 160 \ a_6
          + 176 \ b_6
          \right) \nonumber \\
    & & \hspace*{-1.5cm} 
          + \ \mathcal{O}( \ (d-4)^6 \ ) \ . 
\label{eq:Phi1upto5} 
\end{eqnarray} 
(For the definition of the mathematical constants, see for instance 
Table 1 of~\cite{sun2}). 
The terms up to $(d-4)^3$ agree, within an obvious change of notation, 
with the formula for $J_{11}$ given in \cite{Krakow}; the last two terms, 
of order $(d-4)^4, (d-4)^5$ are new in the literature, and are in full 
agreement with the numerical values for the same quantities given 
in \cite{numcal}. 
\par 
Somewhat surprisingly, the coefficients of the expansion of $\Phi(d,x=0)$ 
in $(d-4)$ at order $n$ are harder to obtain in this way, as they involve 
all the integration constants up to $n$ included. 
\begin{figure}[h] 
\begin{center} 
\includegraphics*[2cm,1cm][14cm,6cm]{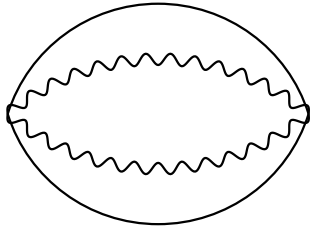} 
\caption{The vacuum 3-loop graph.} 
\end{center} 
\label{fig3} 
\end{figure} 
On the other hand, $\Phi(d,x)$ at $x=0$ corresponds to the vacuum amplitude 
of Fig.(3), which can be easily obtained by direct integration 
of the corresponding Feynman graph amplitude, so that its value can be
used in practice as a check on $\Phi(d,x=0)$.
Conversely, 
by using the explicit value of $\Phi(d,x=0)$ obtained by the direct 
integration, one can fix all the integration constants by imposing the 
proper behaviour at $x=0$ and the value at $x=0$, 
namely $a_0(d)$ in the notation 
of Eq.(\ref{eq:Phiat0}). \par 
For completeness, we list here the exact value of $\Phi(d,0)=a_0(d) $ 
\begin{eqnarray} 
\Phi(d, x = 0 ) & = & \frac{C^{-3}(d)}{(2\pi)^{3(d-2)}} \int 
        \frac{d^dk_1 \ d^dk_2 \ d^dk_3}
             { k_1^2 
              (k_2^2+1)
              (k_3^2+1)
              (k_1 + k_2 + k_3)^2 } \nonumber \\
 &  & \nonumber \\ 
   & = & 
- \ \frac{ 1 }{ 32 \ (d-4)} 
         \frac{  
                \ \Gamma^2( - 1 - (d-4)) \ \Gamma^2(1 + \frac{1}{2} (d-4)) \ 
                \Gamma( - 2 - \frac{3}{2}(d-4)) 
             }{ \Gamma^2(1 - \frac{1}{2}(d-4)) \ \Gamma( - 2 - 2(d-4)) \ 
                \Gamma(2 + \frac{1}{2}(d-4))
              } \ ; 
\label{eq:Phidx0} 
\end{eqnarray} 
its expansion in $(d-4)$ is 
\begin{eqnarray} 
 \Phi(d,x=0) &=& 
       - \frac{1}{24} \frac{1}{(d-4)^3}
       + \frac{7}{96} \frac{1}{(d-4)^2} 
       - \frac{25}{384} \frac{1}{(d-4)} 
       - \frac{5}{1536}
       + \frac{1}{24} \zeta(3)
       \nonumber \\
 &  &
       + (d-4) \left(
          + \frac{959}{6144}
          - \frac{7}{96} \zeta(3)
          + \frac{1}{960} \pi^4
          \right)
       \nonumber \\
 &  &
       + (d-4)^2 \left(
          - \frac{10493}{24576}
          + \frac{25}{384} \zeta(3)
          - \frac{7}{3840} \pi^4
          + \frac{3}{16} \zeta(5)
          \right)
       \nonumber \\
 &  &
       + (d-4)^3 \left(
          + \frac{85175}{98304}
          + \frac{5}{1536} \zeta(3)
          + \frac{5}{3072} \pi^4
          - \frac{21}{64} \zeta(5)
          + \frac{1}{2688} \pi^6
          - \frac{1}{48} \zeta^2(3)
          \right)
       \nonumber \\
 &  &
       + (d-4)^4 \left(
          - \frac{610085}{393216}
          - \frac{959}{6144} \zeta(3)
          + \frac{1}{12288} \pi^4
          + \frac{75}{256} \zeta(5)
          - \frac{1}{1536} \pi^6
          + \frac{7}{192} \zeta^2(3)
         \right. \nonumber \\
   & &
         \qquad \qquad \quad \left.         
          - \frac{1}{960} \pi^4 \zeta(3) 
          + \frac{83}{128} \zeta(7)
          \right)
       \nonumber \\
 &  &
       + (d-4)^5 \left(
          + \frac{4087919}{1572864}
          + \frac{10493}{24576} \zeta(3)
          - \frac{959}{245760} \pi^4
          + \frac{15}{1024} \zeta(5)
          + \frac{25}{43008} \pi^6
         \right. \nonumber \\
   & &
         \qquad \qquad \quad \left.          
          - \frac{25}{768} \zeta^2(3)
          + \frac{7}{3840} \pi^4 \zeta(3) 
          - \frac{581}{512} \zeta(7)
          + \frac{13}{115200} \pi^8
          - \frac{3}{16} \zeta(3) \zeta(5)
          \right)
       \nonumber \\
 &  & + \ \mathcal{O}( \ (d-4)^6 \ ) \ .  
\label{eq:Phidm4x0} 
\end{eqnarray} 
The above expansion corresponds and agrees (again up to trivial changes of 
notation) with the quantities $J_{13}$ of \cite{Krakow} and 
$I_{14}$ of \cite{numcal}. 

\vspace*{1cm}\noindent{\Large {\bf Acknowledgments}} \\ 

\noindent 
We are grateful to J. Vermaseren for his kind assistance in the use 
of the algebra manipulating program {\tt FORM}~\cite{FORM}, by which 
all our calculations were carried out. 

 
\end{document}